\documentclass[prl, twocolumn,showpacs,amsmath,floatfix,superscriptaddress, showkeys]{revtex4-1}
\usepackage{mathrsfs}
\usepackage{graphicx}
\usepackage{dcolumn}
\usepackage{bm}
\usepackage[colorlinks,linkcolor=blue,citecolor=blue]{hyperref}

\usepackage{multirow}
\usepackage{ulem}

\newcommand{\delete}{\bgroup\markoverwith{\textcolor{red}{\rule[0.5ex]{2pt}{1pt}}}\ULon}

\begin{document}
\title{New magicity $N=32$ and $34$ triggered by strong couplings between Dirac inversion partners}
\author{Jia Liu }
\affiliation{School of Nuclear Science and Technology, Lanzhou University, Lanzhou 730000, China}
\affiliation{Key Laboratory of Special Function Materials and Structure Design, Ministry of Education, Lanzhou 730000, China}
\author{Yi Fei Niu }
\affiliation{School of Nuclear Science and Technology, Lanzhou University, Lanzhou 730000, China}
\affiliation{Key Laboratory of Special Function Materials and Structure Design, Ministry of Education, Lanzhou 730000, China}
\author{Wen Hui Long } \email{longwh@lzu.edu.cn}
\affiliation{School of Nuclear Science and Technology, Lanzhou University, Lanzhou 730000, China}
\affiliation{Key Laboratory of Special Function Materials and Structure Design, Ministry of Education, Lanzhou 730000, China}

\begin{abstract}

Inspired by recent experiments, the successive new magicity $N = 32$ and $34$ in Ca isotopes are studied within the relativistic density functional theory. It is illustrated that the strong couplings between the $s_{1/2}$ and neutron ($\nu$) $\nu2p_{1/2}$ orbits, here referred as "Dirac inversion partners" (DIPs), play a key role in opening both subshells $N = 32$ and $34$. Such strong couplings originate from the inversion similarity between the DIPs, that the upper component of the Dirac spinor of one partner shares the same orbital angular momentum as the lower component of the other, and vice versa. Following the revealed mechanism, it is predicted that the magicity $N = 32$ is reserved until $^{48}$S, but vanishes in $^{46}$Si.
\end{abstract}

\maketitle

Atomic nuclei, composed of protons and neutrons, are self-bound quantum many-body system and exhibit typically different magic shells from atoms due to the strong spin-orbital couplings \cite{Mayer1948PR74.235, Haxel1949PR75.1766}. With the worldwide development of radioactive ion beam facilities and detectors, the research area of nuclear physics is largely extended from the stable nuclei to the ones far from the stability, namely the exotic nuclei. Along the extension from the stable to unstable regions, nuclear shell structure can be systematically or even dramatically changed, such as the vanishing of the {traditional magicity} $N=8$ and $20$ in neutron-rich $^{11}$Li \cite{Simon1999PRL83.496} and $^{32}$Mg \cite{Motobayashi1995PLB346.9}, respectively. {On the other hand,} new {magicity} can also arise, for instance the $N=16$ in drip-line magic nucleus $^{24}$O \cite{Ozawa2000PRL84.5493, Hoffman2008PRL100.152502, Kanungo2009PRL102.152501, Tshoo2019PRL109.022501}, which has been reviewed in Ref. \cite{Tanihata2013PPNP68.215}.

In recent years, intensive interests have been attracted on the occurrences of new {magicity} $N=32$ and $34$ in neutron-rich $pf$-shell nuclei. For the magic nature at $N=32$, it has been manifested experimentally by the enhanced $2_1^+$ excitation energy and relatively reduced $B$($E$2; 0$^+$ $\rightarrow$ 2$^+_1$) transition probability in Ar, Ca, Ti and Cr isotopes \cite{Prisciandaro2001PLB510.17, Janssens2002PLB546.55, Dinca2005PRC.71.041302, Gade2006PRC74.021302, Steppenbeck2015PRL114.252501}, and further confirmed by the high-precision mass measurements of exotic Ca and K isotopes \cite{Wienholtz2013Nature498.346, Gallant2012PRL109.032506, Rosenbusch2015PRL114.202501}. More recently, a strong subshell closure at $N = 32$ was also indicated in Sc isotopes through the direct mass measurement, which is found to be quenched in V isotopes \cite{Xu2019PRC99.064303, Reiter2018PRC98.024310}. For the magic nature at $N=34$, it was revealed from the large excitation energy of the $2_1^+$ state ($2043$ keV) in $^{54}$Ca \cite{Steppenbeck2013Nature502.207}.

Inspired by the experimental progresses, lots of theoretical efforts have been devoted to the newly arising magic nature at both $N=32$ and $34$. The large scale shell model calculations with the GXPF1 and KB3G Hamiltonians reproduce the enhanced $2_1^+$ energy at $N=32$ \cite{Janssens2002PLB546.55, Liddick2004PRL92.072502}. Afterwards, the calculations with a beyond-mean-field theory of new generation \cite{Rodriguez2007PRL99.062501} and the {\it ab-initio} one \cite{Hagen2012PRL109.032502} also support the opening of the $N=32$ subshell in Ca isotopes. However, the theoretical discrepancy appears commonly in describing the magic nature at $N=34$, in contrast to the one at $N=32$. For instance, the shell-model calculations using the GXPF1A interaction give a large gap between $\nu$1f$_{5/2}$ and $\nu$2p$_{1/2}$ in $^{54}$Ca, i.e., the subshell $N=34$ \cite{Honma2005EPJA25.499}, while it is not supported by the KB3G interaction \cite{Janssens2002PLB546.55}. Such discrepancy also exists in the {\it ab-initio} calculations. It was pointed out in Ref. \cite{Hergert2014PRC90.041302} that an initial three-body force is necessary to reproduce the shell closures {in} $^{48, 52, 54}$Ca, while a weak shell effect at $N=34$ was presented by the {\it ab-initio} calculations in Ref. \cite{Hagen2012PRL109.032502}. In fact, due to the theoretical discrepancy, one can not help suspecting the robustness of the magic nature at $N=34$.

Until very recently, the first direct mass measurements of $^{55-57}$Ca provide crucial direct evidence for the {magicity $N=34$} in $^{54}$Ca \cite{Michimasa2018PRL121.022506}. It is also worthwhile to mention that the measured $\gamma$-ray spectroscopy of $^{52}$Ar$_{34}$ shows a similar $2_1^+$ energy [1656(18) keV] to that of $^{46}$Ar$_{28}$ [1554(1) keV] \cite{Liu2019PRL122.072502}, an experimental signature of the persistence of the magicity $N=34$ on the proton-deficient side ($Z<20$). These recent experiments make the magicity $N=34$ as robust as the $N=32$ one indeed, thus presenting a challenge for {the theorists} to provide an unified interpretation on the successive new magicity $N= 32$ and $34$ in Ca isotopes, as well as the physics that triggers the magicity $N=34$.

{In this work, the investigations are performed under} the relativistic Hartree and Hartree-Fock approaches \cite{Walecka1974AP83.491, Serot1986ANP16.1}, respectively the relativistic mean field (RMF) and relativistic Hartree-Fock (RHF) models, which own the advantage in deducing the strong spin-orbital coupling naturally. In particular, the important ingredient of nuclear force --- tensor force \cite{Otsuka2005PRL95.232502} can be taken into account naturally via the Fock terms \cite{Jiang2015PRC91.034326, Zong2018CPC42.024101, Wang2018PRC98.034313}. Although many successes have been achieved in describing various nuclear phenomena  \cite{Reinhard1989RPP52.439, Gambhir1990AP198.132, Ring1996PPNP37.193, Bender2003RMP75.121, Vretenar2005PRe409.101, Meng2006PPNP57.470, Liang2015PRe570.1}, it shall be noticed that the popular relativistic Lagrangians, such as the RMF ones DD-ME2 \cite{Lalazissis2005PRC71.024312}  and PK series \cite{Long2004PRC69.034319}, {and the RHF} PKO$i$ ($i=1,2,3$) \cite{Long2006PLB640.150, Long2008EPL82.12001}, fail to reproduce the magic nature at $N=32$ and $34$. Further implemented with the degree of freedom of the $\rho$-tensor coupling,  the  {RHF Lagrangian} PKA1 \cite{Long2007PRC76.034314} brings systematical improvement {in describing nuclear structural properties} \cite{Long2009PLB680.428, Long2010PRC81.031302, Wang2013PRC87.047301}, such as eliminating the spurious shell closures $N/Z = 58$ and $92$ \cite{Long2007PRC76.034314} which commonly exist in the conventional RMF calculations \cite{Geng2006CPL23.1139}. More significantly, PKA1 can reproduce well the successive magicity $N=32$ and $34$ in Ca isotopes \cite{Li2016PLB753.97}, which encourage us to clarify the mechanism of the magicity from the relativistic point of view. 

In this letter, we investigate the new magicity $N=32$ and $34$ in Ca isotopes by using PKA1 \cite{Long2007PRC76.034314}, in comparison with PKO3 \cite{Long2008EPL82.12001} and DD-ME2 \cite{Lalazissis2005PRC71.024312}. Specifically, we concentrate on the physics that triggers the new magicity $N=34$ and further the persistent limit of the magicity $N$=32. For all the calculations with selected Lagrangians, the pairing correlations are considered within the BCS scheme by taking the finite-range Gogny force D1S \cite{Berger1984NPA428.23} as the pairing force. For the isotopes with odd neutron numbers, the blocking effects have been taken into account in the BCS pairing. It shall be remarked that similar results are obtained by the relativistic Hartree-Fock-Bogoliubov theory \cite{Long2010PRC81.024308}. However, it is more straightforward to clarify the mechanism of new magicity under the RHF plus BCS scheme. Besides, the effects of deformation are not considered since most of the concerned nuclei are spherical.

\begin{figure}[htbp]\setlength{\abovecaptionskip}{0.0em}
  \centering
  \includegraphics[width=0.99\linewidth]{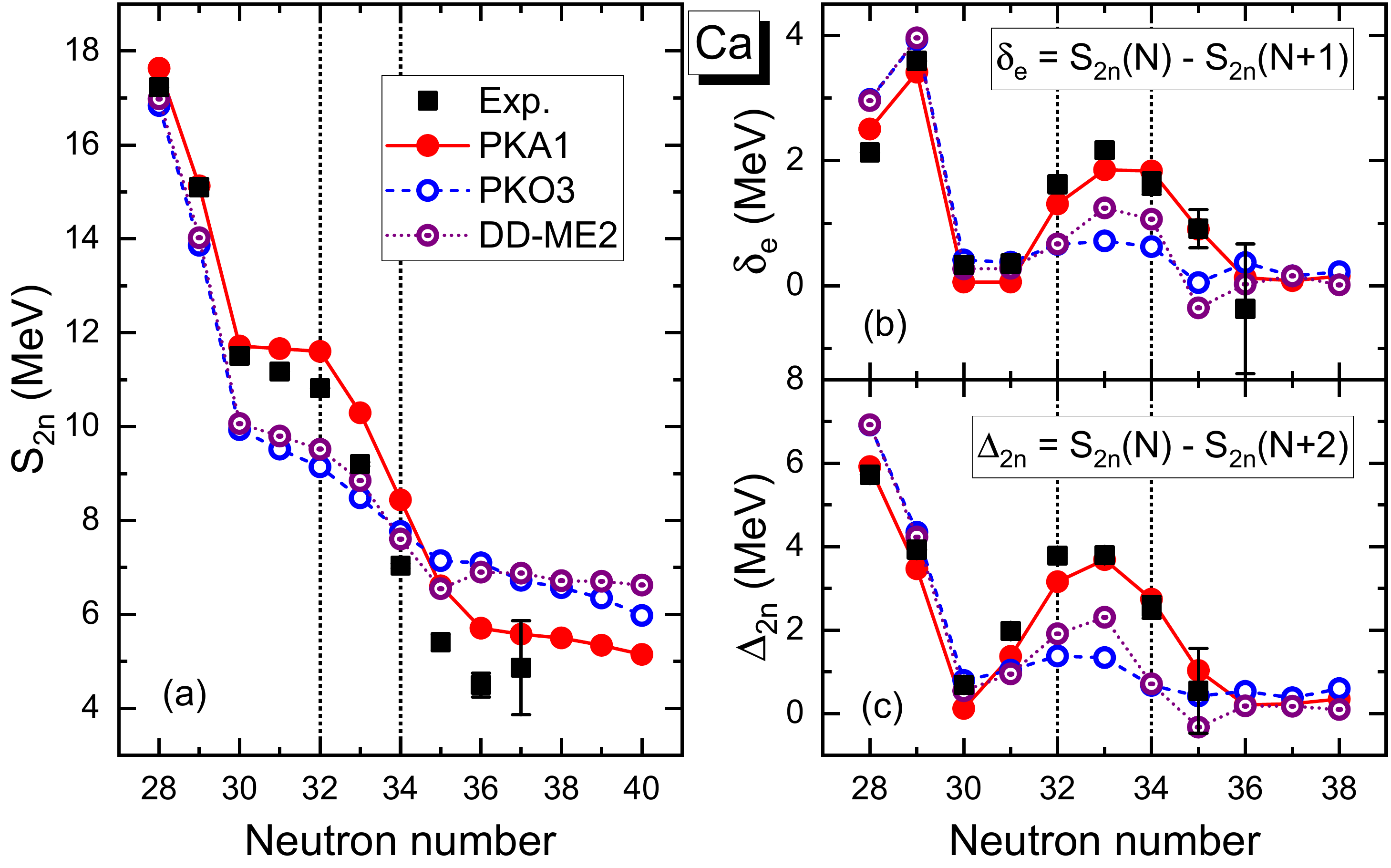}
  \caption{(Color online) Plot (a) shows two-neutron separation energies $S_{2n}$ (MeV) of calcium isotopes, and plots (b) and (c) for the differences, respectively $\delta e = S_{2n}(N)-S_{2n}(N+1)$ and $\Delta_{2n} = S_{2n}(N)-S_{2n}(N+2)$. The results are calculated by PKA1 \cite{Long2007PRC76.034314}, PKO3 \cite{Long2008EPL82.12001} and DD-ME2 \cite{Lalazissis2005PRC71.024312}, as compared to the data \cite{Wang2017CPC41.030003, Michimasa2018PRL121.022506}. }\label{fig:S2n}
\end{figure}

Figure \ref{fig:S2n} (a) shows the {two-neutron separation energies} $S_{2n}$ {(MeV)} calculated by PKA1, PKO3 and DD-ME2, as compared to the experimental data  \cite{Wang2017CPC41.030003} {and very recent measurements} \cite{Michimasa2018PRL121.022506}. {Obviously}, all the selected {models} can properly reproduce the trend of $S_{2n}$ from $N = 28$ to $32$, and PKA1 {presents nice} agreement with the data. After the terrace at $N=30\sim32$, a {significant descending} from $N=32$ to $36$, similar to that from $N=28$ to $30$, is found in the experimental data, {a direct} evidence {of} the successive {magicity} $N=32$ and $34$. Theoretically, the $S_{2n}$ values given by PKO3 and DD-ME2 decrease {from $N=30$} smoothly across $N=32$ and $34$, showing no signal of any shell occurrence. On the contrary, PKA1 properly reproduces the sudden drop at $N=32$ and such drop continues until $N=36$, after which a terrace appears. Although the $S_{2n}$ values are systematically overestimated, the near parallel trend with the data still proves that PKA1 provides appropriate description of the successive {magicity} $N=32$ and $34$. 

From the difference of $S_{2n}$ values of neighboring {isotopes}, namely the filtering function $\delta e = S_{2n}(N)-S_{2n}(N+1)$ {and the two-neutron gap $\Delta_{2n}=S_{2n}(N)-S_{2n}(N+2)$}, one can {quantify the magicity to certain extent} \cite{Satula1998PRL81.3599}. {Figures \ref{fig:S2n} (b) and (c)} show {the $\delta_e$ and $\Delta_{2n}$ values, respectively}. {Similar as the $S_{2n}$ values,} all the selected {Lagrangians} can properly reproduce {both trends of the} $\delta e$ {and $\Delta_{2n}$ values from $N=28$ to $30$}, while the ones at $N=32$ and $34$ are underestimated strikingly by PKO3 and DD-ME2. In contrast to that, nice agreements with the data at both $N=32$ and $34$ are obtained by PKA1, which further prove the reliability of the model in describing the new magicity $N=32$ and $34$. 

\begin{figure}[htbp]\setlength{\abovecaptionskip}{0.0em}
  \centering
  \includegraphics[width=0.99\linewidth]{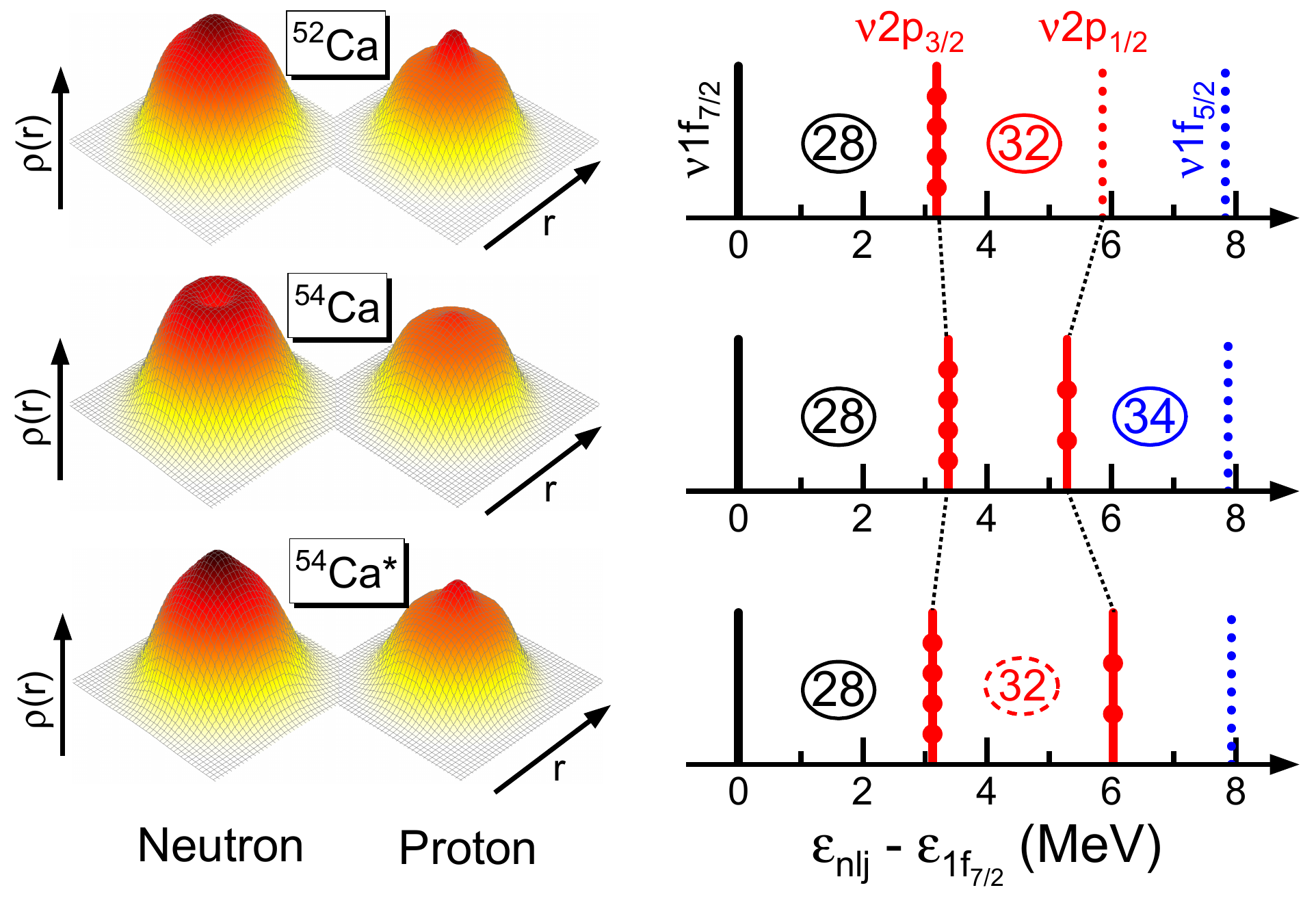}
  \caption{(Color online) Neutron/proton densities (left panels) and neutron ($\nu$) single-particle levels (right panels) for $^{52, 54}$Ca, calculated by PKA1. For the illustration, the last row shows the calculation for $^{54}$Ca that drops the $UL$-terms felt by the $s$-orbits from the DIP $\nu2p_{1/2}$. }\label{fig:density}
\end{figure}

To clarify the mechanism related to the successive magicity $N=32$ and $34$, Fig. \ref{fig:density} shows schematically the 3D plots of both neutron and proton densities (left panels) of $^{52, 54}$Ca and the neutron ($\nu$) single-particle energies $\varepsilon_{\nu nlj}$ (right panels), using the RHF Lagrangian PKA1. At first, let's focus on the first two rows in Fig. \ref{fig:density}. It is seen that both neutron and proton densities {of $^{52}$Ca} show distinct central-bumped structures, being consistent with the large spin-orbit (SO) splitting of $\nu$2p orbits that gives the $N=32$ subshell. This can be interpreted well by the mechanism revealed in Refs. \cite{Rutel2004PRC69.021301, Burgunder2014PRL112.042502, Li2019PLB788.192, Otsuka2018arXiv1805.06501} that the central-bumped or central-depressed density profiles can essentially enlarge or reduce the SO splitting of low-$l$ orbits, such as $p$ states whose probability densities locate at the center of nucleus since the centrifugal repulsion is fairly weak.

While for $^{54}$Ca (the second row in Fig. \ref{fig:density}), the neutron density profile becomes even central-depressed, showing a semi-bubble-like structure, and compared to $^{52}$Ca the central-bumped structure vanishes completely in the proton density. Being consistent with the mechanism mentioned above, the $\nu$2p splitting is notably reduced in $^{54}$Ca. Meanwhile, such dramatic changes of the central densities show little effect on the $\nu1f$ splitting, since their probability densities are driven away from the interior region by strong centrifugal repulsion. Eventually both lead to the occurrence of the $N=34$ {sub}shell in $^{54}$Ca. From the first two rows of Fig. \ref{fig:density}, one can understand the consistent relation between the evolutions of the matter densities and single-particle levels. However, one can't help to ask how the densities can be changed dramatically from the central-bumped structures in $^{52}$Ca to an even central-depressed one in $^{54}$Ca, with only two more neutrons occupying the $\nu$2p$_{1/2}$ orbit.

\begin{figure}[htbp]\setlength{\abovecaptionskip}{0.0em}
  \centering
  \includegraphics[width=0.9\linewidth]{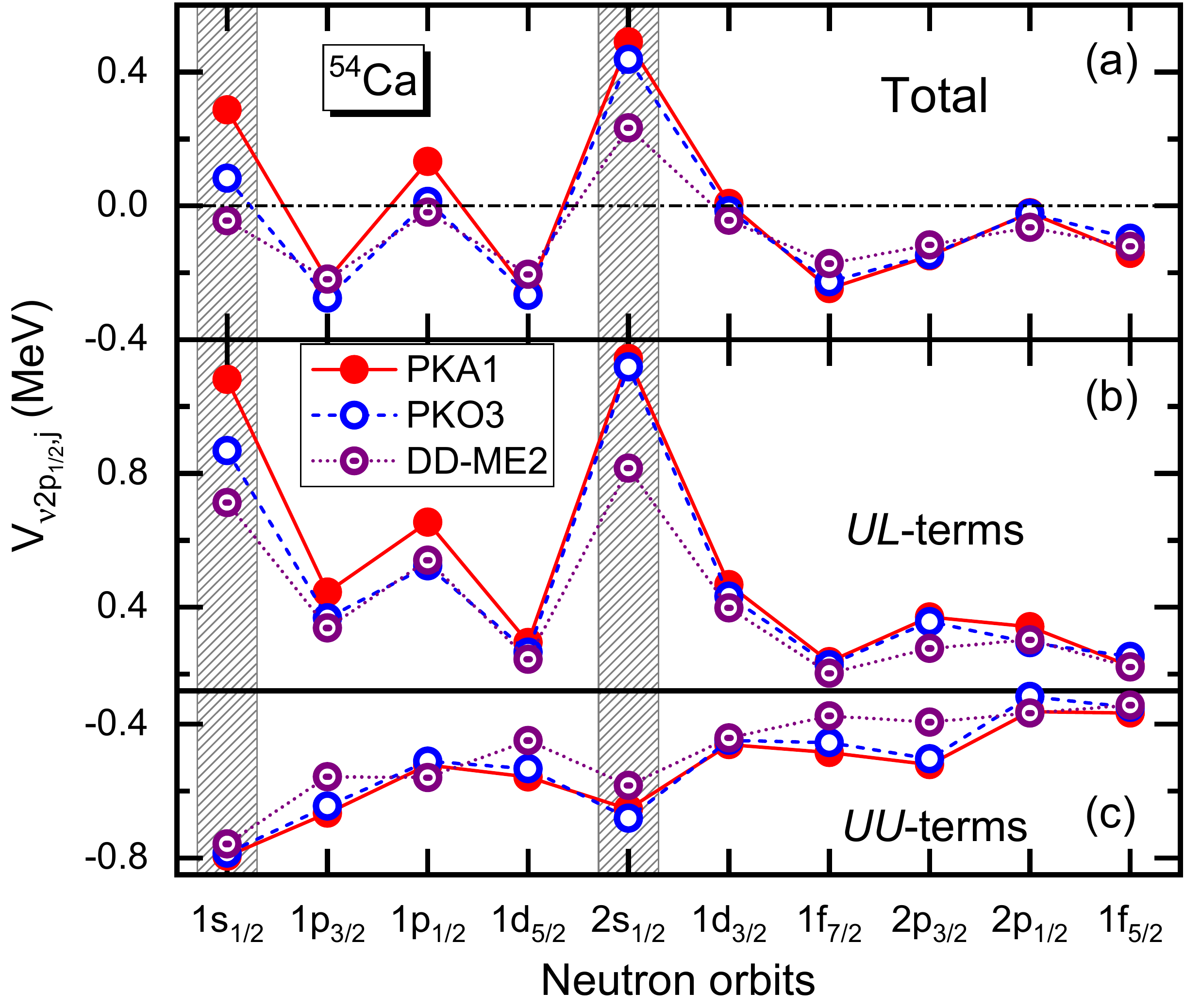}
  \caption{(Color online) Interacting matrix elements [plot (a)] between the neutron $\nu2p_{1/2}$ orbit and the others, namely $V_{2p_{1/2},j}$ (MeV), and the contributions from the $UL$- [plot (b)] and $UU$-terms [plot (c)]. The results are calculated by PKA1, PKO3 and DD-ME2. See the text for details.}\label{fig:inter}
\end{figure}

For further verification, we show the interaction matrix elements between {the} $\nu$2p$_{1/2}$ orbit and the others in Fig. \ref{fig:inter} (a), namely $V_{2p_{1/2},j} = \bar v_{l_1l_2, l_3l_4}$ with $l_1 =l_3 =\nu2p_{1/2}$ and $l_2=l_4 = j$. It is interesting to see that the couplings between the $\nu2p_{1/2}$ and $s$-orbits (emphasized by shadowed area) are repulsive in general, and PKA1 presents the strongest repulsive results. Since from $^{52}$Ca to $^{54}$Ca two extra neutrons occupy the $\nu$2p$_{1/2}$ orbit, such strong repulsion can drive the $s$-orbital neutrons away from the center, as well as the $s$-orbital protons. This results in the reduced central densities in $^{54}$Ca, seeing Fig. \ref{fig:density}, which is in fact due to the reduced $s$-orbital probability densities at the center as compared to $^{52}$Ca.

Under the relativistic scheme, e.g., the RMF and RHF approaches, the Dirac spinors of nucleons contain the upper ($U$) and lower ($L$) components. We notice that the $L$-components of Dirac spinors of the $p_{1/2}$ orbits share the same angular wave functions with the $U$-ones of the $s_{1/2}$ orbits, and vice versa. Here we call such doublet as the "Dirac inversion partners" (DIPs), which are of the same total angular momentum but opposite parity. To understand the repulsive couplings between the DIPs, here $(\nu2p_{1/2}, \nu s_{1/2})$, we decompose the interaction matrix elements $V_{2p_{1/2},j}$ into the $UL$-terms and $UU$-terms as shown in Figs. \ref{fig:inter} (b) and (c), respectively. For the $UU$-term, it contains the contributions only from the $U$-components of Dirac spinors, while the $L$-components contribute the $UL$-terms. Since the $L$-components present tiny contributions to the probability densities,  one would not expect substantial contributions from the $UL$-terms to the interaction matrix elements.

However, from Fig. \ref{fig:inter} (b) it can be seen that the $UL$-terms are in general repulsive and such effects are strongly enhanced for the DIPs $(\nu2p_{1/2},s_{1/2})$, which lead to strong repulsion between the DIPs, although cancelled partly by attractive $UU$-terms in Fig. \ref{fig:inter} (c).
Such enhancement can be understood well from the similarity between the $U$-/$L$-components of $\nu 2p_{1/2}$ orbit and the $L$-/$U$-ones of its DIPs $s_{1/2}$. {Implemented with the} Fock terms, namely the RHF approach, the space parts of the vector couplings, as well as the $\rho$-vector-tensor coupling and the time component of $\rho$-tensor one in PKA1, introduce new couplings between the $U$- and $L$-components of Dirac spinors, which are repulsive but generally missing in RMF. Therefore from RMF to RHF, the $UL$-terms become more repulsive for the DIPs, which indeed enhance the repulsions between the DIPs, seeing Fig. \ref{fig:inter}.

As an implementation, we performed a test calculations with PKA1 for $^{54}$Ca, in which the repulsive $UL$-terms felt by the $s$-orbits from their DIP $\nu2p_{1/2}$ are dropped, and the results are shown in the last row of Fig. \ref{fig:density} (marked as $^{54}$Ca$^\star$). Similar as $^{52}$Ca, the central-bumped structures appear in both neutron and proton densities, and consistently the $\nu2p$ splitting becomes large enough to artificially give the $N=32$ subshell and eliminate the $N=34$ one. Combined with Figs. \ref{fig:density} and \ref{fig:inter}, {it is clear} that the valence neutrons occupying the $\nu2p_{1/2}$ orbit in $^{54}$Ca, via the repulsive couplings with its DIPs $\nu s_{1/2}$ orbits, leads to the dramatic changes of the central densities, which plays an essential role in triggering the magicity $N=34$ of $^{54}$Ca. 

\begin{figure}[htbp]\setlength{\abovecaptionskip}{0.0em}
  \centering
  \includegraphics[width=0.8\linewidth]{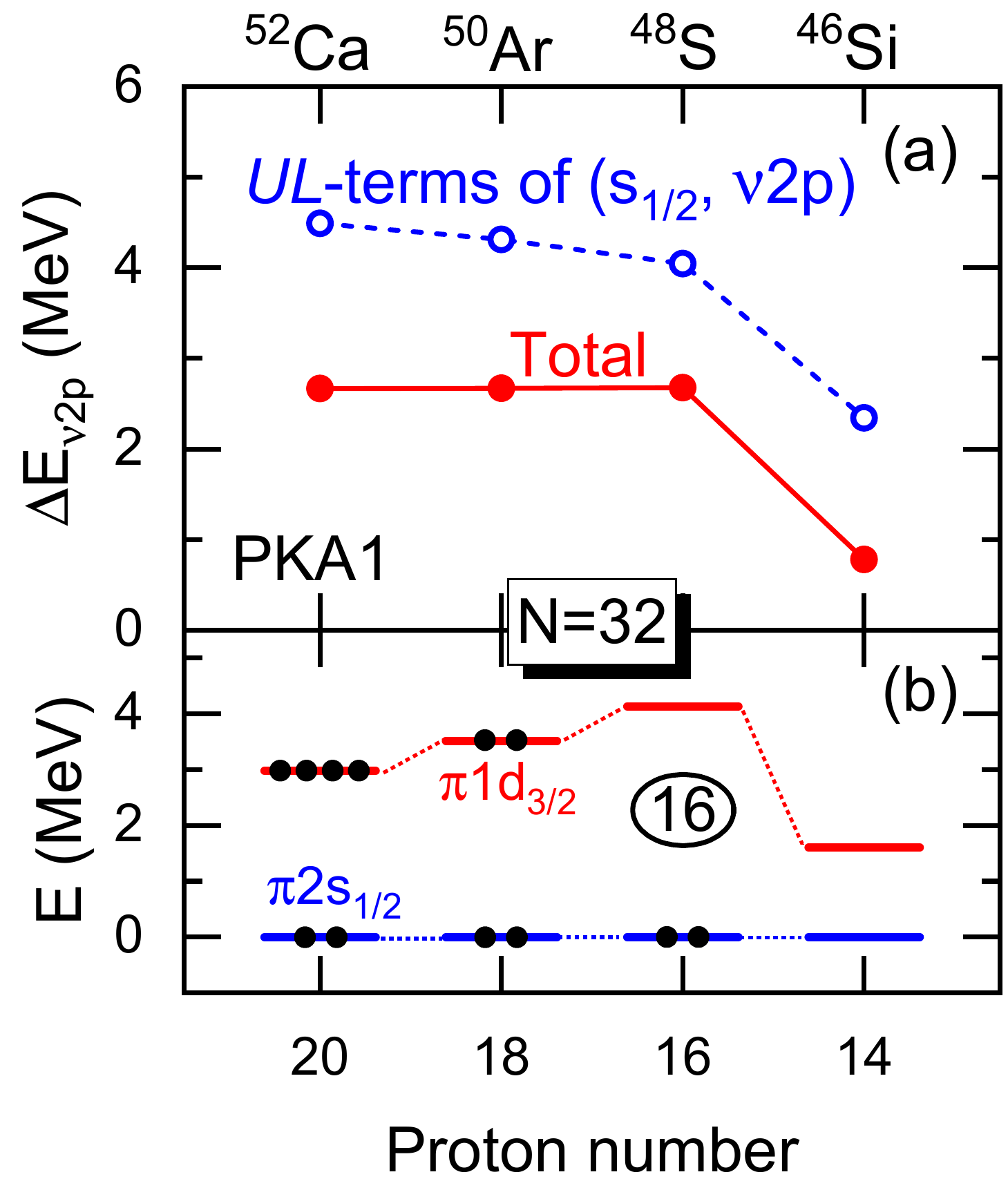}
  \caption{Plot (a) shows the spin-orbit splitting of $\nu$2p states along the isotonic chain of $N=32$ from $^{52}$Ca to $^{46}$Si and the contributions from the $UL$-terms of the couplings between $s_{1/2}$ and $\nu2p$ orbits, calculated by PKA1. Schematically, plot (b) shows the proton orbits $\pi2s_{1/2}$ and $\pi1d_{3/2}$. }\label{fig:48S}
\end{figure}

As indicated from Fig. \ref{fig:inter}, the $UL$-terms of the DIPs ($s_{1/2},\nu2p_{1/2}$) are largely enhanced, compared to the others, e.g., the ones of the ($s_{1/2}, \nu2p_{3/2}$). It is thus appealing to further study the effects in opening the $N=32$ subshell of $^{52}$Ca, i.e., the $\nu2p$ splitting $\Delta E_{\nu2p}$. Figure \ref{fig:48S} (a) shows the evolution of $\Delta E_{\nu2p}$ (MeV) along the isotonic chain of $N=32$ from $^{52}$Ca to $^{46}$Si, as well as the contributions from the $UL$-terms of the couplings between $s_{1/2}$ and $\nu2p$ orbits.
As shown in plot (a), the $\Delta E_{\nu2p}$ value given by PKA1 remains stable from $^{52}$Ca to $^{48}$S and suddenly decreases in $^{46}$Si. Specifically, the $UL$-terms of the ($s_{1/2},\nu2p$) couplings play a dominant role in determining the $\nu2p$ splittings, particularly the evolution.

In fact, such systematics can be interpreted consistently by the valence proton ($\pi$) configurations. From Fig. \ref{fig:48S} (b), the proton spectra of the selected isotones, it is seen that PKA1 presents notable sub-shell gap $Z=16$, particularly in $^{48}$S. It stabilizes the full occupation on the $\pi 2s_{1/2}$ orbit in $^{50}$Ar and $^{48}$S, such that the strong interaction between the DIPs ensures the persistency of the magicity $N=32$ in these two isotones. Further to $^{46}$Si, the empty $\pi2s_{1/2}$ orbit leads to the vanishing of the strong interaction between DIPs $\pi 2s_{1/2}$ and $\nu 2p_{1/2}$, and as a result, the $\nu 2p$ splitting is largely reduced to eliminate the $N=32$ subshell, seeing Fig. \ref{fig:48S} (a). Thus, $^{48}$S is predicted to be the last even isotone that possesses the magicity $N=32$ on the proton-deficient side, which is also a doubly magic nucleus. It is worthwhile to mention that our prediction is consistent with the experimentally revealed magicity $N=32$ in $^{50}$Ar \cite{Steppenbeck2015PRL114.252501}.

In conclusion, the continuous magic nature at $N=32$ and $34$ in Ca isotopes are illustrated by using the relativistic Hartree-Fock (RHF) Lagrangian PKA1 as referred to recent experiments. It is found that large spin-orbit (SO) spliting of $\nu$2p orbits presents the $N=32$ shell in $^{52}$Ca, whereas significantly reduced $\nu$2p splitting, together with nearly unchange $\nu$1f one, leads to the magicity $N=34$ in $^{54}$Ca. Such essential changes of the $\nu$2p splitting can be interpreted self-consistently, following the density evolution from central-bumped structures in $^{52}$Ca to the central-flat (proton) and even cental-depressed(neutron) ones in $^{54}$Ca. It is further illustrated that the dramatic density evolution originates from the strong repulsive interaction between the "Dirac inversion partners" (DIPs) $(\nu2p_{1/2},\nu s_{1/2})$. Finally, we also reveal the mechanism for the appearance of new magicity $N=32$, by analyzing the significant role played by $s_{1/2}$ orbits, which determines the SO spitting of $\nu 2p$ orbits through the strong interaction with its DIP $\nu 2p_{1/2}$.  As a result, $^{48}$S is predicted to be the last even isotone preserving the magicity $N=32$ on the proton-deficient side, and together with the predicted proton subshell $Z=16$, it can be also a doubly magic nucleus.


\section*{Acknowledgement}
This work is partly supported by the National Natural Science Foundation of China under Grant Nos. 11675065, 11875152 and 11711540016, and Fundamental Research Funds for the central universities under Grant No. lzujbky-2019-11. 


\end{document}